# Electronic redistribution around oxygen atoms in silicate melts by *ab initio* molecular dynamics simulation.

R. Vuilleumier[a], N. Sator[b], B. Guillot[b]

[a]  *Laboratoire PASTEUR, UMR CNRS 8640, Département de Chimie, Ecole Normale Supérieure, Paris 75005, France.*

[b] *Laboratoire de Physique Théorique de la Matière Condensée, Université Pierre et Marie Curie (Paris 6), UMR CNRS 7600, case courrier 121, 4 place jussieu,  75252 Paris cedex 05, France.*

The structure around oxygen atoms of four silicate liquids (silica, rhyolite, a model basalt and enstatite) is evaluated by *ab initio* molecular dynamics simulation. Thanks to the use of maximally localized Wannier orbitals to represent the electronic ground state of the simulated system, one is able to quantify the redistribution of electronic density around oxygen atoms as a function of the cationic environment and melt composition. It is shown that the structure of the melt in the immediate vicinity of the oxygen atoms modulates the distribution of the Wannier orbitals associated with oxygen atoms. In particular the evaluation of the distances between the oxygen-core and the orbital Wannier centers and their evolution with the nature of the cation indicates that the Al-O bond in silicate melts is certainly less covalent than the Si-O bond while for the series Mg-O, Ca-O, Na-O and K-O the covalent character of the M-O bond diminishes rapidly to the benefit of the ionic character. Furthermore it is found that the distribution of the oxygen dipole moment coming from the electronic polarization is only weakly dependent on the melt composition, a finding which could explain why some empirical force fields can exhibit a high degree of transferability with melt composition.

# 1. Introduction

Classical molecular dynamics simulations based on empirical interatomic potentials (EPMD) are now widely used to simulate molten silicates of increasing complexity (e.g. [1] and references therein). The advantage of using empirical potentials is that they are easy to implement in simulation codes and are inexpensive in computer time with the result that a large number of atoms (N>1000) can be simulated over long period of time (t>1ns). The disadvantage is that the quality of the results crucially depends on the accuracy of the force field used as input. The advent of the density functional theory (DFT) in quantum chemical calculations [2,3] and its implementation in a molecular dynamics scheme by Car and Parrinello [4] has been a breakthrough for condensed matter physics. However, the very high computational cost (3 to 4 orders of magnitude more expensive in computer time than classical simulations) restricted its use to small systems (N~100 atoms) evolving over short period of time (t~10 ps). With regard to mantle minerals and their melts, the number of studies using *ab initio* molecular dynamics simulations (AIMD) is increasing late years and now cover a rather large range of composition (e.g. $SiO_2$ [5,6], MgO [7,8], $MgSiO_3$ [9,10], $Mg_2SiO_4$ [11-12], $Na_2SiO_4$ [13] and $CaO_{(0.12)}Al_2O_{3(0.21)}SiO2_{(0.67)}$ [14]).

In a recent article [15] we have investigated by AIMD simulation four silicate liquids, representative of the composition range encountered with natural magmas namely, pure silica, an iron-free rhyolite, a model basalt (the anorthite(36wt%)-diopside(64wt%) eutectic composition), and an ultrabasic melt of enstatitic composition ($MgSiO_3$). The structure, ionic diffusivities and infrared absorption spectrum of the simulated melts were compared with data of the literature and with their counterparts evaluated by classical MD simulations using an empirical force field developed recently for multicomponent silicate system [1]. It was shown that the structural parameters of the melts vary little between the two calculations (*ab initio* versus empirical) but the dynamical properties are more sensitive to the details of the interactions. Moreover, it was pointed out that the electronic polarization of the oxygen atoms contributes significantly to the intensity of the infrared absorption spectra of molten silicates.

Thus the electronic structure calculation provided by AIMD is particularly appropriate to get information on the electronic redistribution around atoms on the fly and to shed light on its relationship with the local structure. In evaluating the electronic density around oxygen atoms with the help of the maximally localized Wannier orbitals [16], we show in the following how the electronic density associated with oxygen atoms redistributes as function of the composition and local environment. This finding greatly improves our understanding of the melt structure at the atomic level in pointing out the relationship between the electronic redistribution of oxygen atoms and cationic environment.

## 2. Computational methodology

Ab initio molecular dynamics simulations were performed with the CPMD code [17] using the Kohn-Sham formulation of the density functional theory in the generalized gradient approximation B-LYP [18,19]. Core electrons were not treated explicitely and norm conserving plane wave pseudopotentials of the Trouiller-Martins type were used for all atoms [20] except for aluminium for which a Bachelet-Hamann-Schlüter type pseudopotential was employed [21]. For the cations, Na and K were described by semicore pseudopotentials, while a non-local core correction was used for Mg. All pseudopotentials were transformed to non-local form following the method of Kleinman and Bylander [22], up to $d$ angular momentum for Si, Al and Ca, and $p$ angular momentum for the remaining species. Kohn-Sham orbitals were expanded in plane waves with a kinetic energy cut off of 70 Ry. The Newton equations were solved with a fictitious mass $\mu= 300$ a.u. for the electrons and a time step of 0.048 fs. The simulation cell of constant volume (microcanonical ensemble) was periodically replicated and contained about 96~100 atoms according to the composition of the silicate melt under investigation (see Table 1). The length of the simulation runs were typically 6~7ps, the first 1~2ps of which were discarded.

Since the AIMD calculations are expensive in computer time, it is convenient to optimize the process of equilibration by starting from atomic configurations which are representative of the investigated system. So, the initial configurations for AIMD simulations were prepared from fully equilibrated runs generated by classical MD simulations based upon the empirical potential (EPMD) developed by two of us for silicate melts of various composition [1]. Thus in starting from a configuration generated by EPMD simulation, we have checked that the AIMD calculation (in the microcanonical ensemble) tends to equilibrate rapidly in a few tenths of a ps. The simulations were performed at a high enough temperature (T~2200-2500K) to be sure that we were exploring the liquid phase for all composition. This was verified by analyzing the evolution of the mean square displacements of the atoms with the simulation time. With production runs of only 5ps, it is not possible to check with much accuracy if the imposed density (see below) for the melt under investigation corresponds to P~0 in the simulation. On the other hand it is known that the use of the generalized gradient approximation (GGA) for the exchange correlation energy tends to underestimate by ~2-3% the density of silicate minerals when the use of the local density approximation (LDA), a priori less accurate, leads (fortuitously) to a better estimation [8,10]. In fact GGA remedies some of the weaknesses of LDA, and especially its tendency to overbinding, and the slight underestimation of the density comes from a lack of dispersion energy between atoms [23,24]. In practice the AIMD runs were performed at the experimental density (measured or extrapolated) of the corresponding melt at atmospheric pressure and 2273K namely 2.20 $g/cm^3$ for $SiO_2$ [25], 2.26 $g/cm^3$ for rhyolite [26], 2.55$g/cm^3$ for $an_{36}di_{64}$ [27] and 2.49 $g/cm^3$ for enstatite [28], the temperature of the simulation runs fluctuating around 2360 K for $SiO_2$, 2420 K for rhyolite, 2500 K for $an_{36}di_{64}$ and around 2550 K for molten enstatite.

## 3. Electronic structure of oxygen atoms and relationship with melt structure.

### 3.1 Structure of the silicate melts

The structure of the four simulated melts has been discussed in ref.[15] and we will just briefly summarize here the main conclusions concerning the local environment around oxygen atoms. In silicate systems it is the nature of the oxygen atoms, bridging (BO) or non-bridging (NBO), and their respective population which govern the structure of the melt. The populations of BOs and NBOs are given in Table 2 for the four investigated compositions. Let us recall that an oxygen atom is a BO when it is connected to two network former cations (Si or Al) whereas it is a NBO when it is connected to only one Si or Al. In our calculations two atoms (e.g. Si and O) are connected if their separation is less than the distance associated with the first minimum of the corresponding pair distribution function. In practice this distance is 2.30A for Si-O and 2.55A for Al-O. For the BOs we have distinguished the Si-O-Si, Si-O-Al, and Al-O-Al bonds as also as the triclusters O-(T)$_3$ (with T=Si or Al) which are oxygen atoms shared by three network former cations tetrahedrally coordinated. Moreover, for the NBOs we have distinguished the oxygens linked to one network former cation from the free oxygens only linked to network modifier cations (Mg, Ca, Na or K).

In silica, as expected, all oxygen atoms are BO with a very small proportion of triclusters (0.5%). In the rhyolitic melt although the alkali cations act as charge compensators for Al there is a non negligible population of NBOs (2.5%). Moreover, a small but significant population of oxygen triclusters is predicted (1%). These findings are in agreement with NMR data on charge-balanced aluminosilicate glasses [29-32]. In the basaltic liquid (an$_{36}$di$_{64}$), the degree of depolymerization of the melt is indicated by the high abundance of NBOs (39.1%), while the population of oxygen triclusters is roughly the same as in rhyolite (~1.2%). These results are in agreement with recent findings obtained with high temperature glasses of analog composition [33]. In molten enstatite, the proportion of NBOs is high (65.9%), as expected for a highly depolymerized melt (the standard model of glass structure [34] predicts a value of 66.6% for the ratio $N_{NBO}/O_{Tot}$). However, the NBOs are shared in Si-O-(Mg)$_{n=1,..4}$ structural units, which tend to increase the connectivity of the melt. Incidentally, the amount of free oxygens (oxygen atoms only sharing Mg cations) is low, especially with AIMD, which suggests that there is no (or very little) micro aggregation of $MgO_n$ polyhedra in the melt. Concerning the BOs, approximately 1/3 of them are linking two $SiO_4$ units exclusively, while the remaining 2/3 are shared by two $SiO_4$ units and 1- or 2-$MgO_{n=4,5,6}$ polyhedra, the consequence being an increase of the connectivity of the melt. Nevertheless, this network is very fragile at liquid temperature owing to the relative weakness of the Mg-O interaction with respect to the Si-O one (the large difference of viscosity between this ultrabasic melt and a silica melt is the consequence of this feature).

## 4. Electronic redistribution around oxygen atoms using localized Wannier orbitals.

The great advantage of the electronic structure calculation (AIMD) is its ability to give information about the electronic redistribution around atoms induced by the interactions within the melt. However the main challenge in a first-principles theory is to define relevant atomic or molecular surfaces that would enable one to partition the electron density between the various atoms or molecules present in the system. Various solutions to this molecular boundary problem have been proposed since the pioneering population analysis by Mulliken [35,36]. In the case of finite systems with periodic boundary conditions, as used here, Vanderbilt and King-Smith [37] and Marzari and Vanderbilt [38] have proposed a very efficient method for generating maximally localized orbitals from Wannier functions developed originally for periodic systems [16]. In practice, maximally localized Wannier orbitals are obtained through an appropriate unitary transformation (accounting for periodic boundary conditions) of the Kohn-Sham orbitals evaluated during the AIMD run (maximally localized orbitals imply that an appropriate functional is minimized to minimize the spread of the orbitals, for details see [38]). In our AIMD calculation where the pseudopotential approximation is used, only electrons around oxygen atoms are considered since the cations are simply described by ionic cores (except for Na and K which are defined with semicores). For each oxygen atom, 4 doubly occupied Wannier orbitals are considered (8 electrons, the total charge of the ion being -2e), the 2 additional oxygen-core electrons being not involved in the bonding properties. The Wannier orbitals (WO) and the Wannier orbital centers (WOC) were evaluated every 100 steps during the AIMD run (for a definition of a WOC see [39,40]).

For the four melts investigated here, the four WOs are on average tetrahedrally distributed around each oxygen atom (for an illustration see Fig.1), the width of the angular distribution being equal to ~20° in silica and ~23° in rhyolite, $an_{36}di_{64}$ and enstatite. The distribution of the intra atomic distances between the oxygen core and the four WOCs was evaluated by taking into account all the oxygen atoms (BO and NBO) of the simulated melt and by averaging over the AIMD steps for which the WOs were evaluated. The result is shown in Fig.2 where the distribution of core-WOC distances is found to be bimodal with one maximum (centered about 0.44 A) which corresponds to an electronic displacement induced by Si-O-Si interactions, whereas the maximum centered about 0.29 A in silica and rhyolite (and ~0.34 A in $an_{36}di_{64}$ and enstatite) corresponds to lone pair orbitals (lp) interacting eventually with network modifying cations (Ca, Mg, Na and K). These assignments can be further quantified by evaluating the core-WOC distribution associated with BOs and NBOs.

Because there are only BOs in silica (see Table 2), the two distances 0.29 and 0.44 A can be used as benchmarks for non interacting lp orbitals and Si-O bonding orbitals, respectively. Thus, in rhyolite (see Fig.3) the oxygens linked to two Si atoms (Si-O-Si) exhibit the same core-WOC distribution than BOs in pure silica, as expected. For oxygens interacting simultaneously with Si and Al atoms (Si-O-Al), a new feature appears around 0.39 A in the distribution (see Fig.3) which is assigned to Al-O interactions, while the peak corresponding to lp orbitals is slightly shifted toward larger distances by the presence of K (from 0.29 to 0.30A) and Na cations (see the shoulder around 0.33 A), as the latter act as charge compensating cations for Al.

In molten $an_{36}di_{64}$ (see Fig.4) the distribution of core-WOCs distances associated with BOs involved in Si-O-Si interactions is similar to what it is observed in silica and rhyolite, except that a significant broadening of the lp peak located at r~0.29 A is observed because of the presence of a population of network modifiers (Ca and Mg) in the environment of the lp orbitals. Furthermore, for BOs involved in Si-O-Si, Si-O-Al and Al-O-Al interactions, the signatures of Si-O and Al-O interactions are clearly visible around 0.43 and 0.39 A, respectively. In the case of Al-O-Si and Al-O-Al interactions, the peak associated with free lp orbitals collapses and is replaced by a broad band peaking around 0.34 A that corresponds to $Ca-O_{lp}$ and $Mg-O_{lp}$ interactions, although it is not easy to discriminate between these two cations. Correspondingly, in the case of NBOs, the distribution of core-WCOs distances is dominated by a strong peak about 0.34 A (see –O-Si in Fig.4), the peak associated with Si-O interactions (r~0.44 A) being much less intense than for BOs.

In molten enstatite a pattern consistent with the above findings is observed (see Fig.5). Thus when a NBO is shared between one Si and one Mg atom (see Mg-O-Si in Fig.5), one WOC is pulled out by the bonding with Si (r~0.45 A), another WOC is pulled out more moderately by the interaction with Mg (r~0.34 A) while the remaining two WOs are kept almost free of interaction (r~0.30 A). Furthermore, when a NBO is shared between one Si and two Mg atoms (see $Mg_2$-O-Si in Fig.5), only the distances $r_{Si-O}$=0.45 A and $r_{Mg-O}$=0.34 A are clearly visible on the distribution whereas the peak at $r_{lp}$~0.29 A corresponding to lp orbitals is virtually absent. An explanation could be that the two Mg atoms are in a bifurcated position with respect to three Wannier orbitals. In contrast, when a bridging oxygen is shared between two Si and one Mg atom (see Mg-O-$Si_2$ in Fig.5), not only the peak at 0.44 A is well developed as in the case of rhyolite or silica, a feature which is expected for a Si-O-Si bridging oxygen, but the lp peak at 0.29 A is also present which means that a WO is kept free. As for the fourth WO, it is involved in an Mg-O bond characterized by a core-WOC distance slightly shorter than the one occurring in the case of a NBO (e.g. r~0.32 A instead of ~0.345 A for Mg-O-$Si_n$ with n=1-3, see Fig.5). This finding could suggest that the relatively large displacements of the WOCs involved in Si-O interactions limit somewhat the displacements of the lp orbitals when they interact with other cations.

At this stage it is worth noting that a related analysis based on AIMD calculations has been reported for the behavior of the water molecule in the liquid phase by Silvestrelli and Parrinello [41]. For the isolated water molecule, the four WOCs associated with the oxygen atom are tetrahedrally oriented and their distance from the oxygen core is 0.53 A for the two OH covalent bonds and 0.30 A for the lone pair orbitals. In liquid water every water molecule is hydrogen bonded to on average approximately four neighbors, the OH bonds being proton donors and the lone pair proton acceptor. The result is that the tetrahedral hydrogen bond network tends to pull out the lone pair of each molecule from 0.30 A to 0.33 A (because the lone pair interacts with protons of nearby molecules) whereas the WOCs located on the OH intramolecular bonds draw nearer to the oxygen core from 0.53 A to 0.50 A (due to the slight repulsion between the lone pair of the proton acceptor molecule and the WO of the proton donor molecule). By analogy, in liquid silica the Si-O bond plays the role of the OH covalent bonds but the lone pair on the oxygen atoms interacts more weakly with its environment in silica than in liquid water. As a matter of fact, the core-WOC distance is about 0.29 A for the lone pair in silica as compared with 0.33 A in liquid water. Concerning the Si-O interaction, the distance of 0.44 A found here in silicate melts is not very far from the value of 0.50 A found for the OH covalent bond in liquid water, a similarity which advocates the importance of the covalence in the Si-O bond (for a discussion about the long standing debate about the nature of the Si-O bond see [42-45]). In this context, our findings about the core-WOC distances in silicate melts suggest that the Al-O bond is certainly much less covalent than the Si-O bond (r~0.39 A as compared with 0.44 A), and that for the series Mg-O, Ca-O, Na-O and K-O (r~0.34 - 0.33 and 0.30 A, respectively) the covalent character of the interaction diminishes drastically to the benefit of the ionic character. Of course our purpose is not to settle the important question concerning the ionic/covalent nature of the cation-oxygen interaction in silicates but only to give some new clues.

In a previous article [15] we emphasized the contribution of the electronic polarization of oxygen atoms to the intensity of the IR absorption spectrum of silicate melts. To better characterize the polarization of the electronic clouds associated with oxygen atoms in response to the field generated by the melt structure, we have evaluated from the calculated core-WOC distances, the dipole moment carried by the oxygen atoms into the melt. The distribution of the modulus of the oxygen dipole moment averaged over all BOs and NBOs is shown in Fig.6 for the four compositions (see the curve labeled BO+NBO in Fig.6). The most probable value of the dipole moment (~2.5 D) is essentially composition independent but the more depolymerized the melt the narrower the distribution. In fact, the distribution of the dipole moment averaged over all oxygen atoms does not change drastically with composition. This may help to understand why a rigid ion model like the one recently proposed [1] exhibits an appreciable transferability with melt composition although it ignores the electronic polarization effects, the latter ones being approximately accounted for in an effective way through the parameters of the ion-ion interactions. Nevertheless our calculations show that the dipole moment distributions associated with BOs and NBOs are not identical to each other (see Fig.6) and these differences induce subtle modification of the oxygen dipole moment distribution with composition. More precisely, NBOs tend to sample larger values of the dipole moment than BOs do. This tendency is all the stronger that the network modifier cation located in the immediate vicinity of a NBO is of low field strength (in Fig.6 compare the curves labeled NBO for rhyolite, where $Na^+$ and $K^+$ are the network modifier cations, with $an_{36}di_{64}$ and enstatite where the network modifier cations are $Ca^{2+}$ and $Mg^{2+}$).

To make the link between the oxygen dipole moment distribution and the structure of the melt we propose a very simple model which shows that the oxygen dipole moment is governed essentially by the T-O-T bond angle distribution (where T = Si or Al). Let us consider a BO linked to two Si atoms. It is assumed that on average two WOCs lie in the Si-O-Si plane when the two others (the lone pair) are located in the perpendicular plane (for an illustration see Fig.1). Our model assumes that the displacement of the WOCs along the axes of the tetrahedron is governed solely by the Si-O-Si bond angle (i.e. $r_{WOC}\sim f(\theta_{SiOSi})$), the tetrahedral configuration of the four Wannier orbitals remaining unchanged (the slight dispersion around tetrahedrality which is evaluated about ±5° is neglected here). Of course the analytic expression of $f(\theta_{SiOSi})$ is not known a priori but it can be shown that with any reasonable guess for this function one can reproduce satisfactorily both the distribution of the core-WOC distances and the distribution of the oxygen dipole moment in using the Si-O-Si bond angle distribution calculated by AIMD (see Fig.10 in [15]). For illustration the distribution of the modulus of the dipole moment associated with BOs linked to two Si atoms is presented in Fig.7, assuming the following simple relationship for the displacement of the WOCs along the tetrahedral axes,

$$r_{\parallel,\perp} = a_{\parallel,\perp} \sin(\theta-\theta_T)/2 + b_{\parallel,\perp} \cos(\theta-\theta_T)/2 \qquad (9)$$

where $r_{\parallel,\perp}$ are the distances between the oxygen core and the WOCs lying into the plane ($r_{\parallel}$) defined by the angle $\theta_{Si-O-Si}$, or perpendicular to it ($r_{\perp}$), $\theta_T$ is the tetrahedral angle (109.5°) and where $a_{\parallel,\perp}$ and $b_{\parallel,\perp}$ are elongation parameters fitted to best reproduce (not shown) the core-WOC distributions presented in Figs 2-5. The resulting dipole moment is evaluated by simple algebra in considering that each Wannier orbital carries a charge -2e and that the ionic core (charge +6e) is the origin of the coordinates. The agreement between the distribution of the oxygen dipole moment calculated by AIMD and the one predicted by the model in using the simulated bond angle distribution [15] is remarkable considering the simplicity of the model (notice that the same level of agreement can be obtained with NBOs). An important result is that the maximum of the dipole moment distribution corresponds to the maximum of the bond angle distribution (in Fig.7 the arrow indicates the value of the dipole moment at the maximum of the bond angle distribution). It means that the electronic polarization of the oxygen atom is essentially driven by the local structure.

**5. Conclusion**

Ab initio molecular dynamics simulations have been performed to investigate the electronic redistribution around oxygen atoms in four silicate liquids covering a range of composition representative of the variety of natural magmas. Through the use of maximally localized Wannier orbitals the present ab initio calculations allow a detailed analysis of the electronic structure of the oxygen atoms as function of the local cationic environment and melt composition. It is shown that the distribution of Wannier orbitals around the oxygen atoms is governed by the local structure of the melt. In particular the evaluation of the distances between the oxygen-core and the orbital Wannier centers and their evolution with the nature of the neighboring cations indicates that the Al-O bond in silicate melts is certainly less covalent than the Si-O bond while for the series Mg-O, Ca-O, Na-O and K-O the covalent character of the M-O bond diminishes rapidly to the benefit of the ionic character. Furthermore it is found that the distribution of the oxygen dipole moment coming from the electronic polarization is only weakly dependent on the melt composition, a finding which could explain why some empirical force fields can exhibit a high degree of transferability with melt composition. From a more general point of view, the results presented here can be useful to better understand NMR, IR and Raman spectroscopic data and can help to improve the empirical force fields implemented into MD simulation codes for silicate melts (for recent advances see [46]).

## BIBILIOGRAPHY

**Table 1**

Chemical composition (in weight fraction) of the silicate melts simulated in this study. In parenthesis are the numbers of cations of each species used in the simulations of the corresponding melt.

| | SiO$_2$(wt%) | Al$_2$O$_3$(wt%) | MgO(wt%) | CaO(wt%) | Na$_2$O(wt%) | K$_2$O(wt%) |
|---|---|---|---|---|---|---|
| silica | 100.0 (33) | 0.0 (0) | 0.0 (0) | 0.0 (0) | 0.0 (0) | 0.0 (0) |
| rhyolite | 77.9 (26) | 12.7 (5) | 0.0 (0) | 0.0 (0) | 4.6 (3) | 4.7 (2) |
| an$_{36}$di$_{64}$ | 50.7 (18) | 14.3 (6) | 11.3 (6) | 23.7 (9) | 0.0 (0) | 0.0 (0) |
| enstatite | 59.9 (20) | 0.0 (0) | 40.1 (20) | 0.0 (0) | 0.0 (0) | 0.0 (0) |

**Table 2**

Populations (in %) of bridging oxygens (BO) and non-bridging oxygens (NBO) in the simulated melts. Notice that $O-Si_3$, $Si_2-O-Al$, $Al_2-O-Si$ and $O-Al_3$ are triclusters.

| | BO | | | | | | | | NBO | | | |
| | Si-O-Si | Si-O-Al | Al-O-Al | $O-Si_3$ | $Si_2-O-Al$ | $Al_2-O-Si$ | $O-Al_3$ | $\sum = N_{BO}$ | Si-O | Al-O | free oxygens | $\sum = N_{NBO}$ |
|---|---|---|---|---|---|---|---|---|---|---|---|---|
| silica | 99.5 | | | 0.5 | | | | 100. | | | | |
| rhyolite | 67.1 | 27.8 | 1.5 | 0.3 | 0.4 | 0.2 | 0.1 | 97.4 | 2.5 | 0.0 | 0.0 | 2.5 |
| an36di64 | 24.0 | 33.0 | 2.5 | 0.0 | 0.9 | 0.3 | 0.0 | 60.7 | 36.1 | 3.0 | 0.0 | 39.1 |
| enstatite | 34.0 | | | 0.0 | | | | 34.0 | 65.5 | | 0.4 | 65.9 |

**Figures**

**Fig.1** Two Wannier orbitals around a bridging oxygen (Si-O-Si) in rhyolite (color in line). (a) Wannier orbital corresponding to a lone pair and (b) Wannier orbital along the O-Si bond. The volumes delimited by the isocontours contain 1.5e.

**Fig.2** Distribution function of core-Wannier orbital center distances ($R_{C\text{-}WOC}$) associated with oxygen atoms in the four simulated melts.

**Fig.3** As in Fig.2 but for bridging oxygens (Si-O-Si and Si-O-Al) in rhyolite. The arrows indicate the ionic species interacting with the corresponding Wannier orbital. Notice that in the case of Si-O-Al bridging oxygens the lone pair orbitals are preferentially linked to the charge compensating cations Na and K.

**Fig.4** As in Fig.3 but for bridging (Si-O-Si, Al-O-Al and Si-O-Al) and non-bridging (-O-Si) oxygens in $an_{36}di_{64}$.

**Fig.5** As in Fig.3 but for bridging (Si-O-Si, Mg-O-Si$_2$) and non-bridging (Mg-O-Si, Mg$_2$-O-Si) oxygens in enstatite.

**Fig.6** Distribution function of the modulus of the oxygen dipole moment averaged over all oxygen atoms (BO+NBO), or over BO and NBO atoms, respectively. In the case of silica the very tiny difference between the Si-O-Si curve and the curve labeled (BO+NBO) is due to oxygen triclusters (see Table 2).

**Fig.7** Distribution function of the modulus of the dipole moment associated with BO atoms (Si-O-Si): full curves, simulation results; dotted curves, our model based upon the bond angle distribution (see text). The arrows indicate the maximum of the Si-O-Si bond angle distribution for each composition.

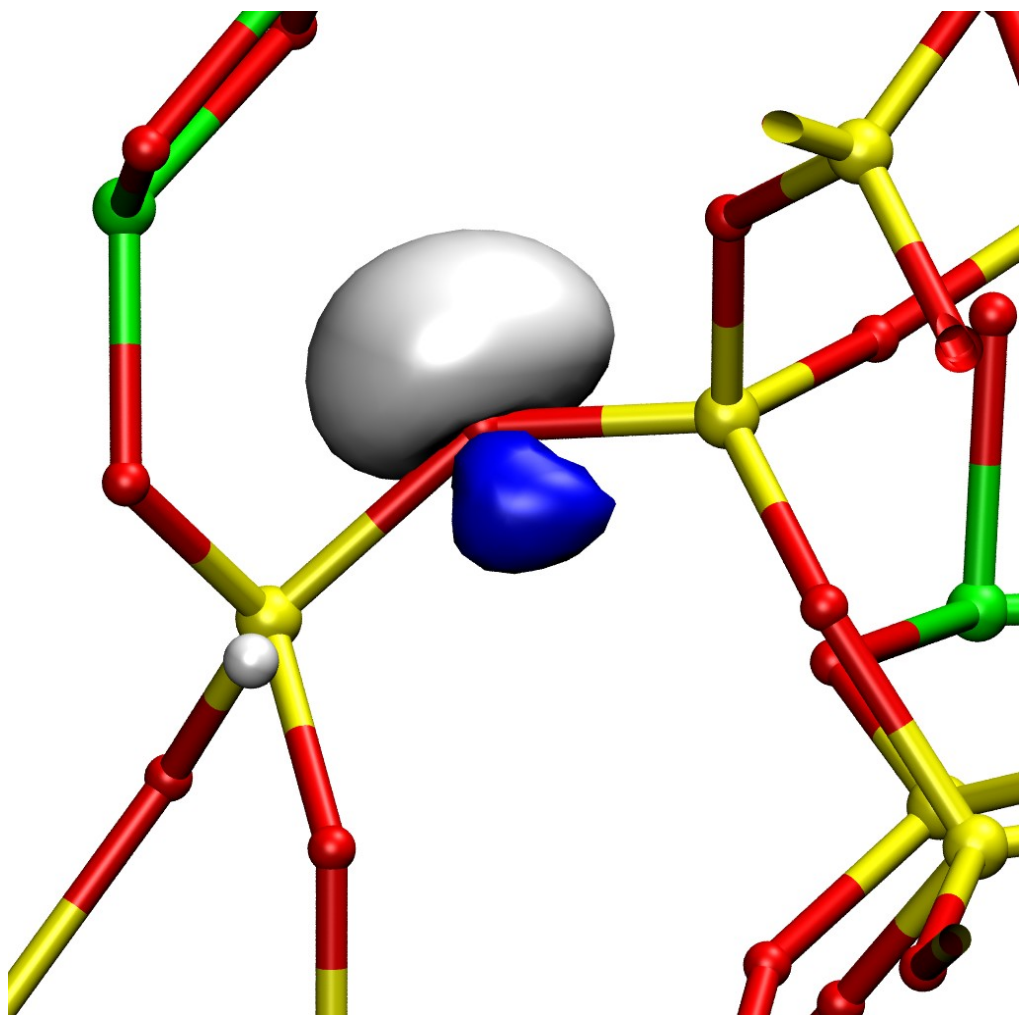

**Fig.1a**

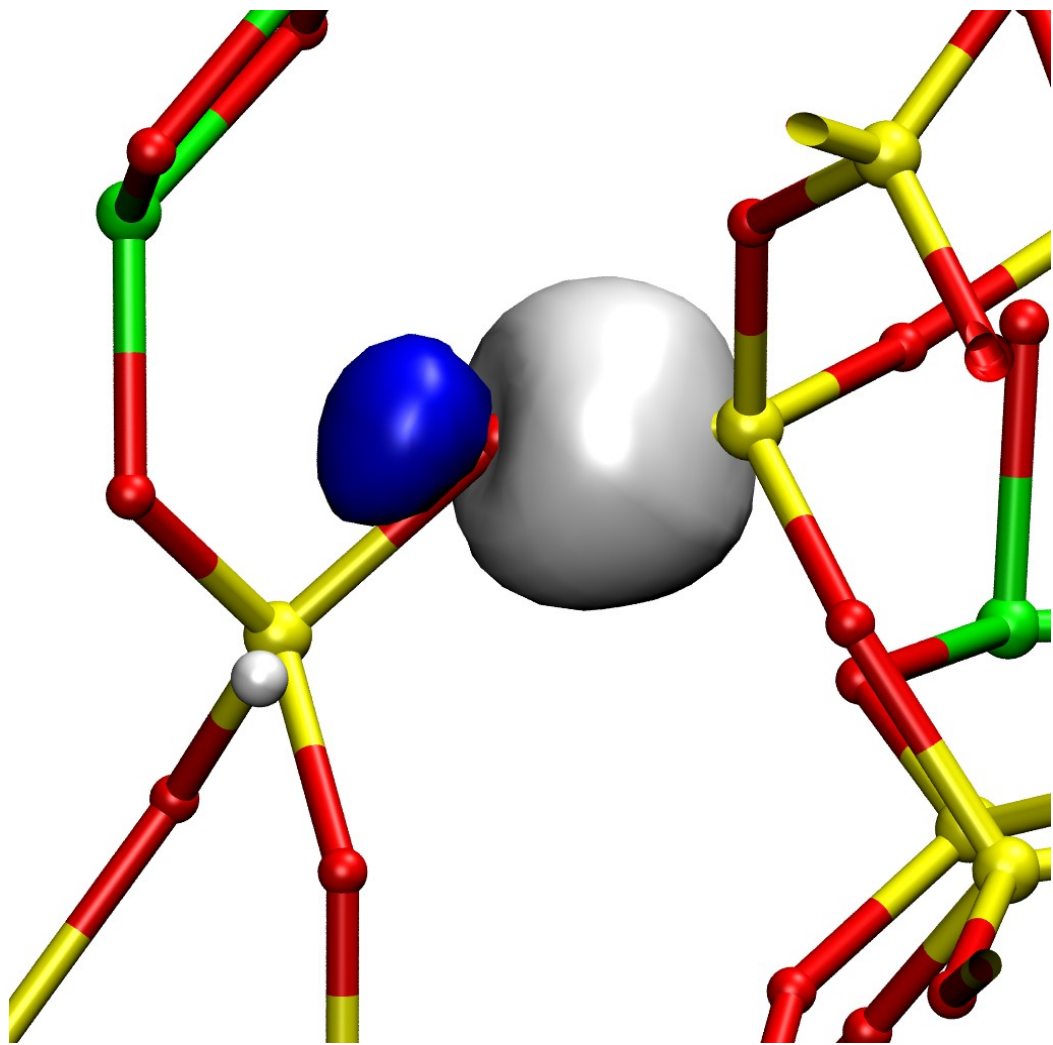

**Fig.1b**

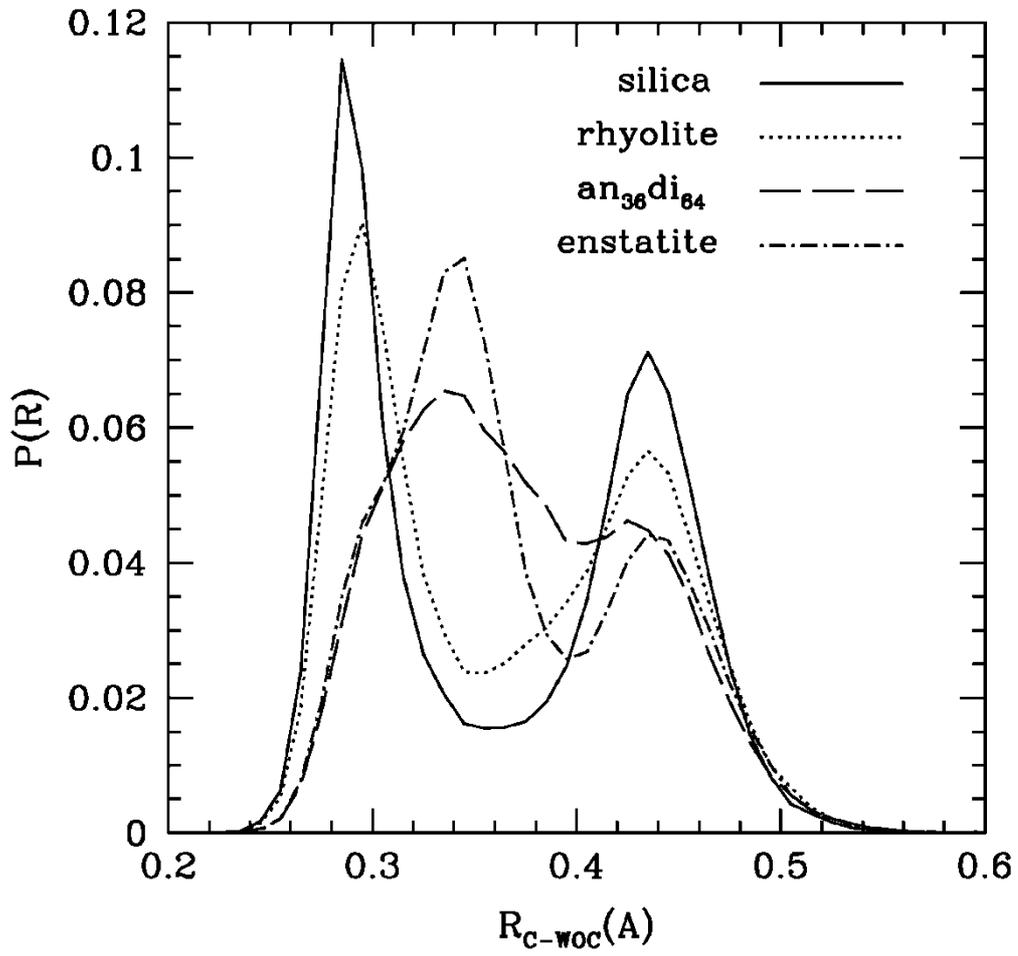

**Fig.2**

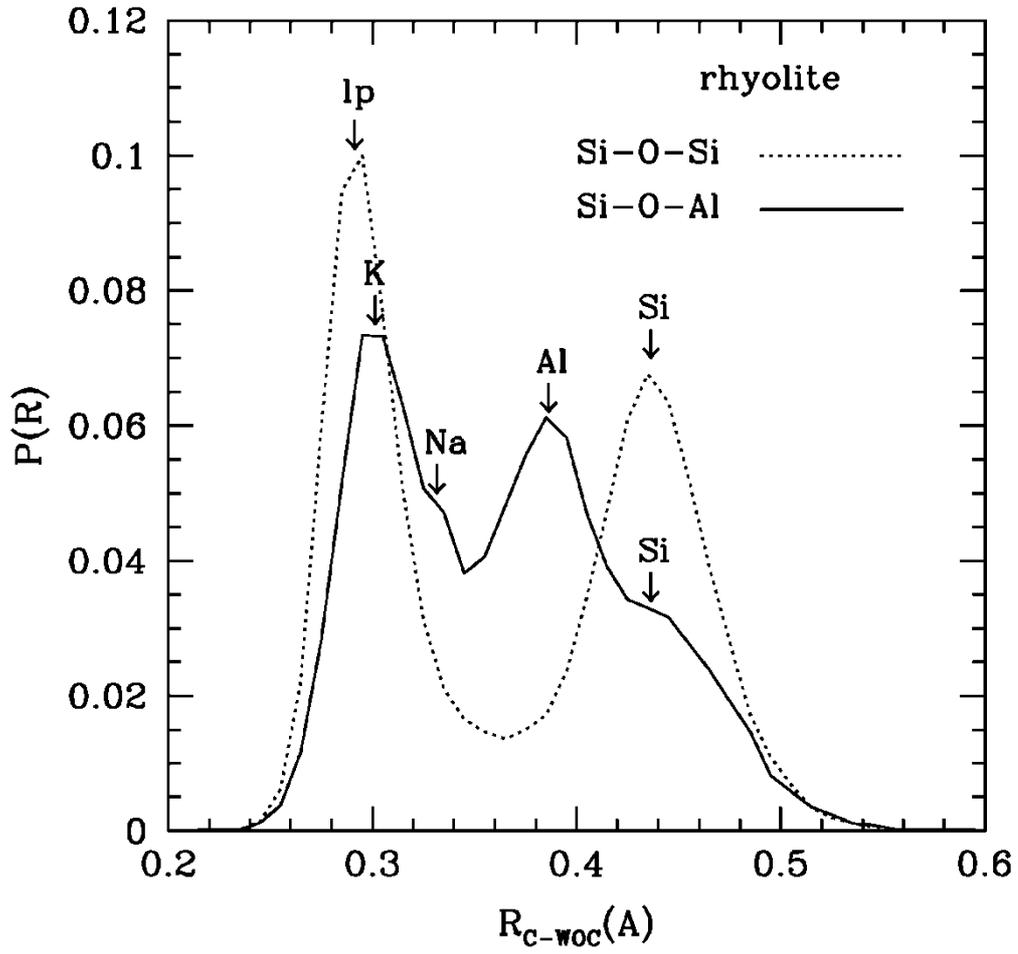

**Fig.3**

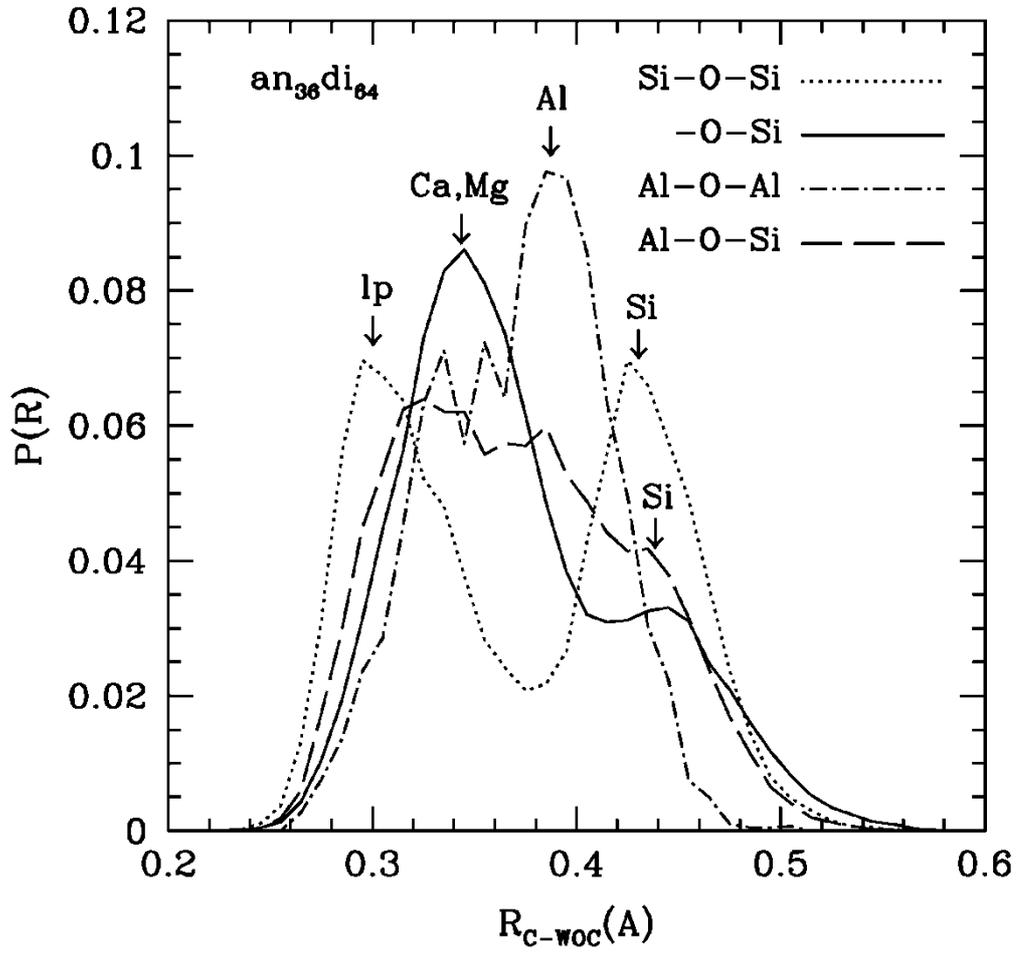

**Fig.4**

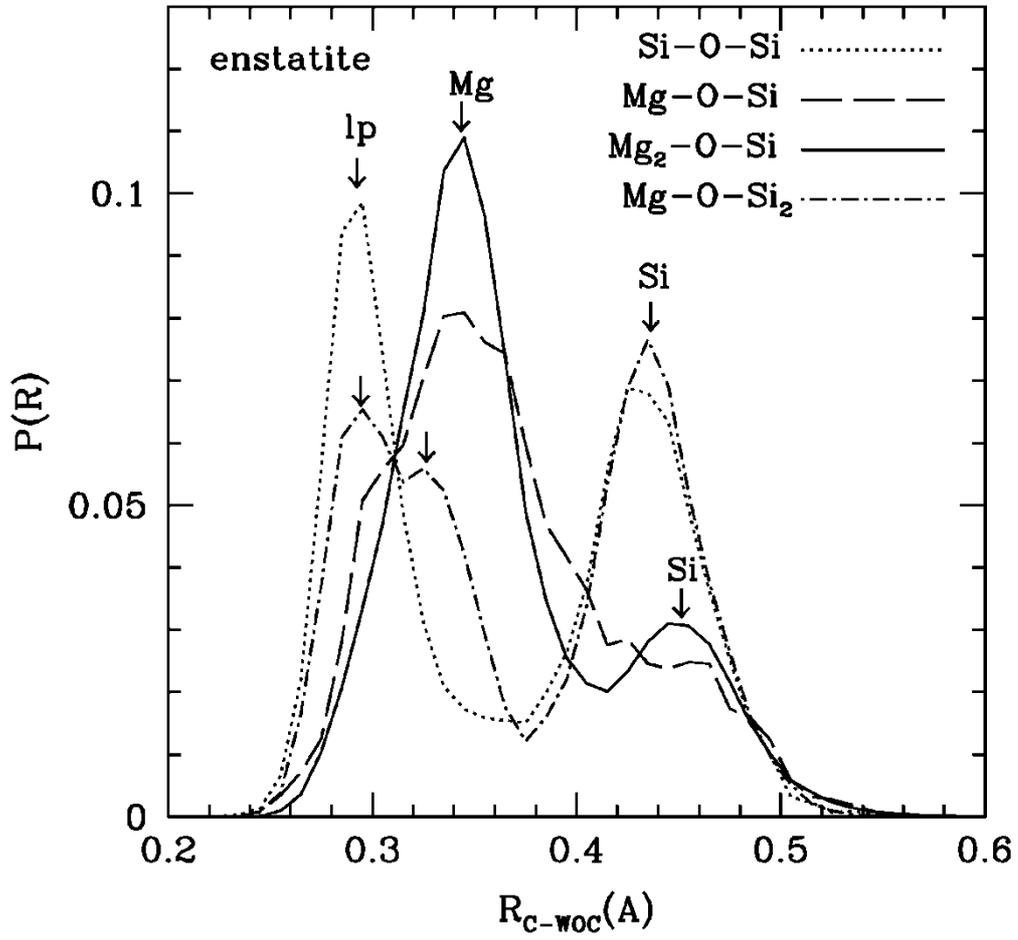

**Fig.5**

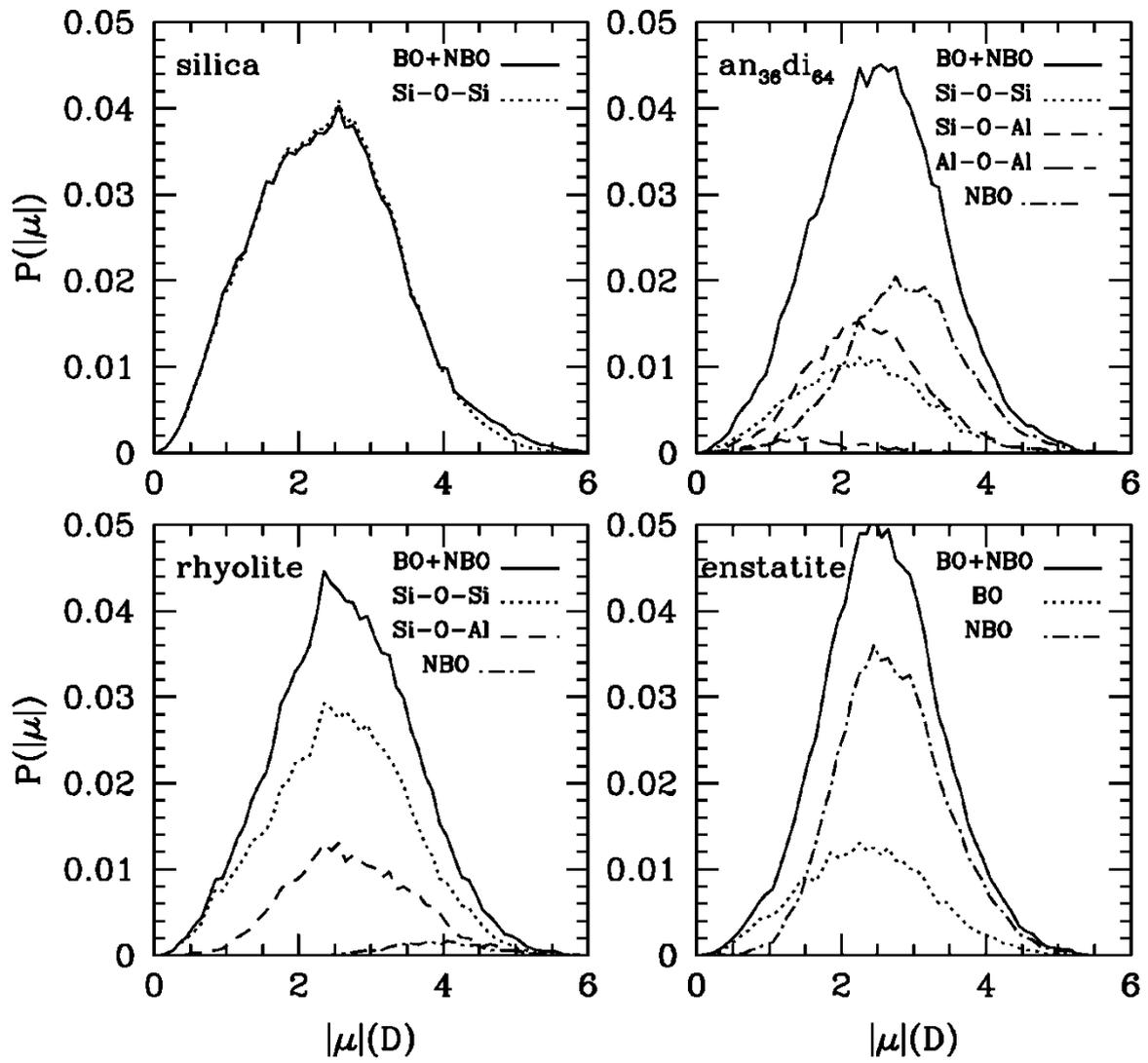

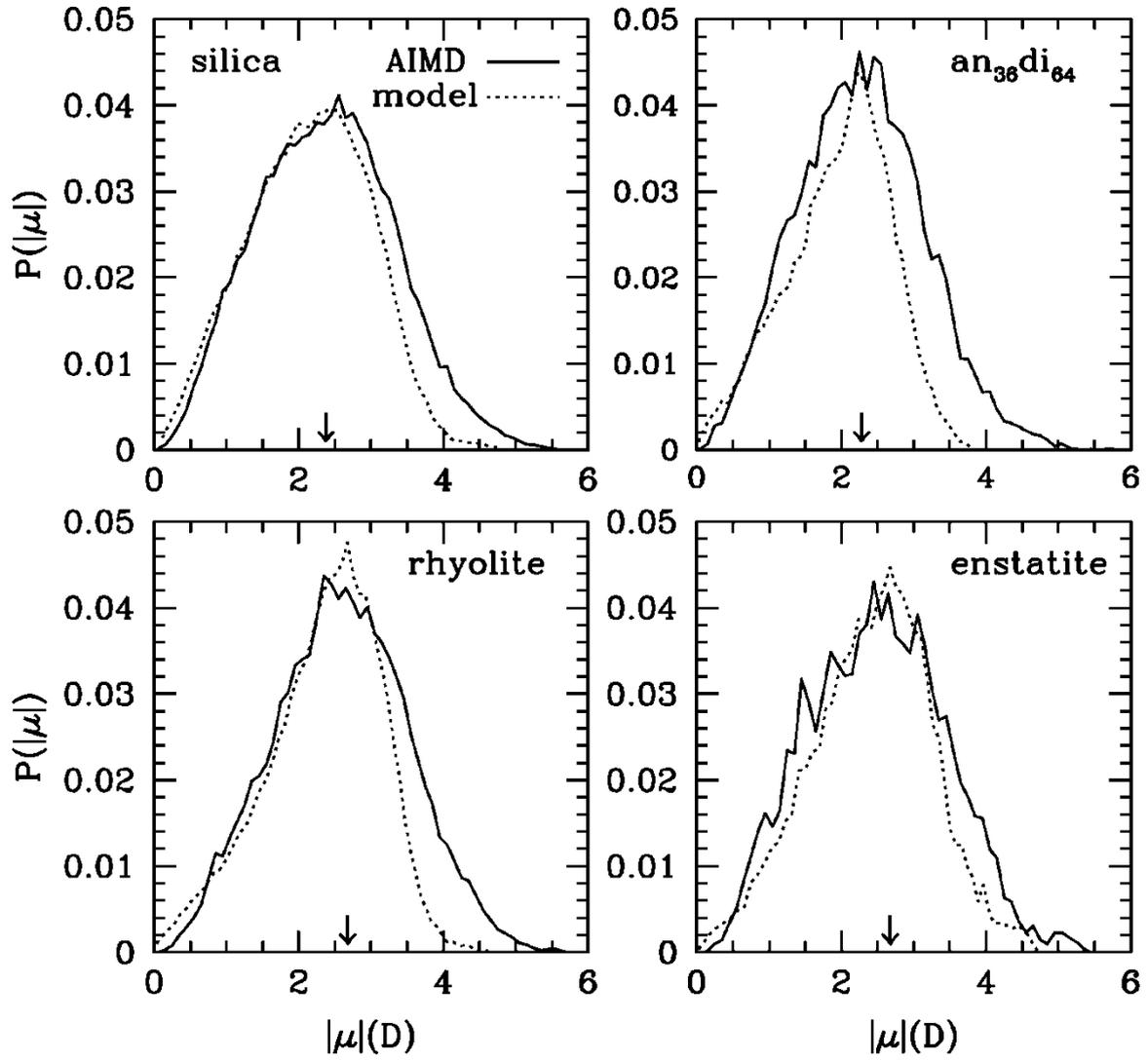

**Fig.7**